\documentclass{article}

\usepackage{arxiv}
\usepackage{bm}
\usepackage[round]{natbib}  
\usepackage[utf8]{inputenc} 
\usepackage[T1]{fontenc}    
\usepackage{hyperref}       
\usepackage{url}            
\usepackage{booktabs}       
\usepackage{amsfonts}       
\usepackage{nicefrac}       
\usepackage{microtype}      
\usepackage{lipsum}
\usepackage{graphicx}
\usepackage{amsmath}
\graphicspath{ {./images/} }

\title{Narrative Sharpens Gender Gaps: Surveying Film Characters with LLM Agents}

\author{
Vivienne Bihe Chi \\
 Computer and Information Science\\
  University of Pennsylvania\\
  Philadelphia, PA, USA \\
  \texttt{vchi@seas.upenn.edu} \\
   \And
Reyhan Jamalova\\
 Computer and Information Science\\
  University of Pennsylvania\\
  Philadelphia, PA, USA \\
  \texttt{jreyhan@seas.upenn.edu} \\
  \And
  Lyle Ungar \\
  Computer and Information Science\\
  University of Pennsylvania\\
    Philadelphia, PA, USA \\
  \texttt{ungar@cis.upenn.edu} \\
  \And
 Sharath Chandra Guntuku\\
Computer and Information Science\\
  University of Pennsylvania\\
   Philadelphia, PA, USA \\
  \texttt{sharathg@seas.upenn.edu} \\
}

\usepackage{bm}
\begin{document}
\maketitle
\begin{abstract}
Mainstream film is one of the richest sources of cultural content that AI systems learn from. 
Yet we have few tools for measuring  the gender values it encodes. We present a proof-of-concept framework that turns fictional film characters into surveyable LLM agents. 
Using 160 U.S.\ films (1990--2019), we build 734 character agents from script dialogue and scene descriptions, condense their personas via expert-style reflections, and simulate World Values Survey gender-attitude responses. 
Agents reproduce systematic gender differences without explicit demographic prompting, suggesting attitudes emerge from behavior rather than identity labels. 
Benchmarked against historical survey data, agents exaggerate gender gaps and show greater decade-to-decade volatility than real populations. 
Narrative sharpens rather than homogenizes gender contrasts, complicating the consistent-input assumption underlying cultivation theory's mainstreaming mechanism. 
AI systems trained on such corpora may inherit this stylization before any model-level amplification occurs.
\end{abstract}


\section{Introduction}
Mainstream film is one of the most widely shared upstream sources
of the gender norms that play out in online communities and
collaborative platforms today. 
Cultivation theory has long held that sustained exposure to consistent media messages shapes how
audiences understand social reality, making certain attitudes feel
natural and widely shared~\citep{gerbner1976}.
For researchers
interested in where gender dynamics in collaborative settings come
from, film is an obvious place to look. The difficulty is
measurement.

Prior computational work has taken steps in this direction.
Working within this venue, \citet{jang2019} used
automated image analysis to quantify visual gender bias in
commercial films, measuring screen time, spatial framing, and
on-screen presence across a large corpus. Related work has
analyzed gender patterns in film dialogue and tracked how gendered
language associations shift over decades~\citep{Martinez2022-rc, Hamilton2016-bx,Bamman2013LearningLP}. These methods are powerful, but they
measure representation and language rather than expressed
attitudes. They tell us who appears and how they speak, not what
values the narrative asks audiences to recognize as plausible or
legitimate.

Recent advances in LLM-based simulation suggest a different
approach. 
\citet{Farrell2025}
argue that LLMs function less like minds and more like cultural
repositories: systems that loosely replicate, summarize, and
transmit collective human cultural information, comparable in kind
to libraries or the printing press. From this view, mainstream
film is one of the most concentrated tributaries feeding that
repository. Auditing what gender values film contains is therefore
a prerequisite for understanding what AI systems trained on
narrative data inherit. LLM-based character agents, conditioned on
script-derived memories, offer a practical instrument for that
audit. Agents grounded in personal text have been shown to
reproduce survey responses and behavioral patterns of real
individuals~\citep{park2024, Argyle2023}, suggesting the
same approach can surface the attitudinal content of fictional
characters at scale.


We build 734 character agents from 160 U.S.\ films released between
1990 and 2019. Each agent is grounded in dialogue and scene
descriptions drawn from the script, condensed into structured
persona reflections by three expert-role LLM personas representing
psychology, linguistics, and sociology. We then prompt these
agents to answer three gender-attitude items from the World Values
Survey (WVS) \citep{wvs} and compare their responses to historical U.S. survey
data from the same period. 

We treat this comparison as a measurement instrument rather than a
validity test. Prior work on synthetic survey respondents has
asked whether LLM-simulated responses are accurate enough to
substitute for human data~\citep{bisbee2024, santurkar2023opinionslanguagemodelsreflect}. We
ask a different question: what does systematic divergence from
population baselines reveal about the source corpus? When the gap
is patterned and theorizable, it becomes signal rather than noise.


Our results bear this out. Character agents reproduce gendered
attitudinal patterns consistent with how gender is portrayed in
film. Compared to actual WVS respondents, however, they exaggerate
gender gaps and show more variation across decades than real
populations do. 
Cultivation theory's 'mainstreaming' mechanism assumes consistent encoded inputs~\citep{gerbner1976}; our findings complicate that assumption at the character level, where gender contrasts are already sharpened before any audience exposure occurs.
AI systems trained on such corpora may inherit these sharpened contrasts before any model-level amplification takes place.

\section{Methods}
In this study, we compare gender attitudes in the United States from 1990–2019 by aligning two sources of evidence: survey responses from real-world respondents and simulated responses from fictional film characters from the same period.
\subsection{Survey Data: Gender Attitudes in World Values Survey}
The World Values Survey (WVS) \citep{wvs} is a large-scale cross-national survey program that measures public values and social attitudes. Since 1981, WVS has conducted nationally representative surveys in waves approximately every five years. For the United States, we use responses from waves 2 through 7, covering the period 1990–2019.

To measure gender attitudes, we focus on three survey items repeatedly fielded across these waves:  
\begin{enumerate}
    \item  ``When jobs are scarce, men should have more right to a job than women,''  
    \item  ``On the whole, men make better political leaders than women do,'' and  
    \item ``A university education is more important for a boy than for a girl.''  
\end{enumerate}

Respondents expressed endorsement of these male-privileging gender norms on 5-point ordinal scales. 
We aggregate responses by gender and decade to enable comparison with simulated responses from film characters.

\subsection{Film Character Corpus}

To construct a corpus of fictional characters for agent simulation, we use the MovieSum dataset \citep{moviesum}, which contains structured movie scripts including dialogue and scene descriptions. We enrich this dataset with film metadata from the Open Movie Database (OMDb) API \citep{omdbapi}, including release year, genre, and cast information.

For this proof-of-concept study, we sample U.S. films released between 1990 and 2019. Films are sampled uniformly by decade (1990s, 2000s, 2010s) and stratified across major genre categories available in OMDb (e.g., drama, comedy, action) to ensure broad coverage of mainstream film narratives.

Lead characters are identified using the primary credited actors listed in the film metadata. We use actor gender and age at the time of release as proxies for character gender and age,  following prior computational film work~\citep{Bamman2024,Martinez2022-rc}. While imperfect, these proxies have proven sufficient for population-level gender comparisons, and we apply it under the same conditions here.

From each script, we extract all dialogue lines spoken by each lead character as well as scene descriptions referencing that character. These elements correspond to the ``dialogue'' and ``scene description'' tags in the formatted movie scripts in the MovieSum dataset. The resulting corpus contains 734 character agents (250 female and 484 male) across 160 films. On average, each character has 180 dialogue lines and 133 action descriptions, which form the narrative evidence used to construct agent memories.

\subsection{Character Agent Memory Construction}

We represent each character as a generative agent adapted from the generative-agent architecture introduced in \citet{park2024}. Each character agent stores both structured metadata and a memory bank derived from the screenplay. The metadata includes the character’s name, gender, estimated age, and the film’s release year (used as the agent’s time period). We use release year as a pragmatic temporal proxy, acknowledging that diegetic setting may differ for period films. However, with a corpus median of 75,832 IMDb user votes, these films plausibly reflect values that large contemporary audiences found culturally legible or aspirational rather than strict period-accurate beliefs. The memory bank consists of narrative observations extracted from the script. Each memory node corresponds to either (1) a line spoken by the character or (2) a scene description mentioning the character and an action they perform.
This memory structure allows the agent to ground its responses in the narrative evidence present in the screenplay while preserving the chronological record of the character’s behavior and dialogue.

\subsection{Persona Condensation via Expert Reflections}

Because the raw memory banks can contain hundreds of observations per character and vary widely in length, we introduce a condensation step to summarize the character’s persona.
We employ three LLM-based expert personas representing different disciplinary perspectives: psychology, linguistics, and sociology. Each expert model receives the full memory bank for a character and generates five evidence-based reflections describing the character’s traits, motivations, social roles, and implied value orientations. Reflections are generated using a fixed prompt template and low sampling temperature (0.1) to encourage consistent outputs.
This process yields 15 structured reflections per character. These reflections serve as a condensed representation of the character’s behavioral tendencies and social positioning within the narrative. 
Because these reflections aggregate repeated behaviors and social interactions, they provide a higher-level representation of value-relevant patterns that may not be directly observable from individual lines of dialogue.
For example, a reflection might note that a character ``manages high-stakes interactions (e.g., organizing logistics, vetting participants, and exercising authority) in male-dominated settings.'' Such reflections are grounded in specific dialogue and actions, providing interpretable links between narrative evidence and inferred values.
In subsequent steps, they provide the primary conditioning context used to simulate survey responses.
Overall, the pipeline proceeds in three stages: (i) extraction of character-level narrative evidence from scripts, (ii) condensation into structured persona reflections, and (iii) simulation of survey responses conditioned on these reflections. This design separates raw narrative evidence from higher-level persona abstraction, allowing the model to reason over aggregated behavioral patterns rather than isolated events.

\subsection{Character Agent Survey Emulation}

We simulate survey responses by prompting each character agent to answer the same gender-attitude questions used in the WVS\footnote{See Appendix~\ref{app:prompts} for the survey emulation prompt template.}. Each prompt includes the character's metadata (name, age, and time period) and the set of expert-generated reflections summarizing the character's persona.
Each agent answers the three WVS gender-attitude items using the original ordinal response scale. Responses are generated using the GPT-5-mini model \citep{openai2025gpt5mini} with a structured prompt instructing the model to respond from the perspective of the character.
Crucially, the prompt does not include explicit gender instructions; gender differentiation, if it appears, must emerge from the narrative evidence encoded in the persona reflections. This design follows Hall's (\citeyear{hall1980}) distinction between encoded and decoded meaning: the agents recover the preferred reading the script constructs, not the interpretation any particular viewer might bring to it.

\begin{figure*}[t]
\centering
\includegraphics[width=\linewidth]{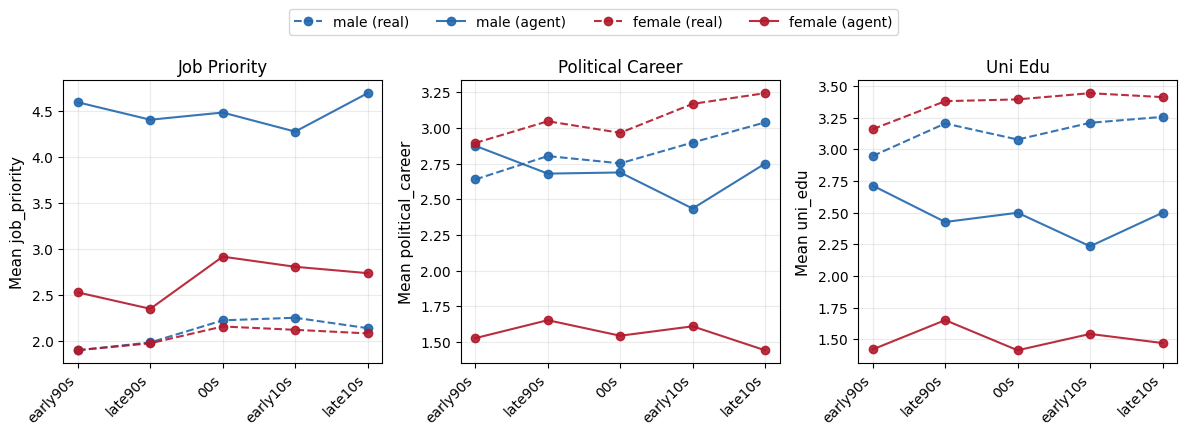}
\caption{Mean responses by decade comparing agent-simulated data with real U.S. World Values Survey data for survey items ``job priority for men'', ``men make better political leaders'', and ``university education more important for boys''. Dashed lines represent real human survey responses (male: blue, female: red), while solid lines represent agent-generated responses. Across topics, real responses remain relatively stable with small gender gaps, whereas agent responses display larger fluctuations and stronger gender differences. This pattern suggests that narrative-based reconstructions encode more polarized and variable attitudes than those observed in real populations.}
\label{fig:image}
\end{figure*}
\section{Results}
The analysis highlights two primary findings. First, within the reconstructed character population we observe systematic gender differentiation: female character agents disagree more strongly than male characters with male-privileging statements about job priority (``When jobs are scarce, men should have more right to a job than women''; Welch's $t = 11.17$, $p < .001$), political leadership (``On the whole, men make better political leaders than women do''; Welch's $t = 15.64 $, $p < .001$), and university education (``A university education is more important for a boy than for a girl ''; Welch's $t = 14.81 $, $p < .001$). This suggests that the script-grounded reflections encode sufficient social and relational context for gendered attitudinal differences to emerge from narrative evidence rather than explicit demographic prompting. Robustness checks using Mann-Whitney U tests yield the same pattern of significance across all three items (all $p < .001$). Because characters are nested within films, both tests treat observations as independent and likely underestimate uncertainty; these results should be interpreted as indicative pending mixed-effects replication.
We do not detect reliable directional trends across age or decade within the agent responses;
greater decade-to-decade variability reported below reflects higher variance around a flat trend, not a consistent temporal trajectory.

When compared against the World Values Survey as a diagnostic benchmark, the agents diverge from historical population patterns in a consistent direction. 
Cultivation theory's mainstreaming mechanism assumes consistent encoded inputs~\citep{gerbner1976}; at the character level, we find the opposite pattern, where gender contrasts are already sharpened before any audience exposure occurs, complicating that assumption.
In the real U.S. data, gender differences are generally modest and relatively stable across decades.
As shown in Figure \ref{fig:image}, agent responses exhibit substantially greater variability across decades than the relatively stable trends observed in survey data, suggesting that reconstructed attitudes are sensitive to narrative composition and character sampling rather than reflecting gradual population-level change.
Supporting this pattern, when comparing simulated to real responses across (gender × decade) cells, simulated means are systematically lower (more gender-egalitarian) than real means for two items: ``men make better political leaders'' ( $\Delta M_{\,\text{sim-real}} = -0.77$; $t = -3.50$, $p < 0.01$) and ``university education is more important for a boy'' ($\Delta M_{\,\text{sim-real}} = -1.21$; $t = -6.13$, $p <0.001$). For ``job priority for men,'' the simulated survey response mean is higher than the real survey response means  ($\Delta M_{\,\text{sim-real}} = 1.57$; $t = 5.42$, $p <0.001$).
Taken together, these discrepancies indicate that while agents generate coherent and internally consistent attitudes, they amplify gender differentiation relative to observed population baselines.

\section{Discussion}

Our results support three connected findings.
\paragraph{Gender emerges from behavior, not labels.}
Gendered attitudinal structure appeared without any explicit
demographic prompting. We did not instruct agents to respond as
gendered actors; gender differentiation arose from dialogue and
action. This is consistent with West and
Zimmerman's~\citep{west1987} account of gender as something
accomplished through situated interaction rather than a fixed
property of a person. Scripts are doing gender on every page, and
the pipeline recovers it from behavior rather than identity
labels.

\paragraph{Narrative sharpens rather than homogenizes.}
The divergence from WVS baselines is not random. Agents exaggerate
gender gaps and show greater decade-to-decade volatility than real
survey respondents. 
As argued above, if cultivation's mainstreaming mechanism depends on consistent encoded inputs~\citep{gerbner1976}, the polarization we observe at the character level complicates that assumption:
at the level of individual character portrayal, encoded gender attitudes are already polarized and volatile before any audience receives them. If the inputs to cultivation are systematically sharpened rather than homogeneous, the mechanism the theory relies on operates on more heterogeneous material than the framework assumes.
Narrative polarizes gender attitudes rather than smoothing them. This makes sense given how stories work. Protagonists are written to embody aspirational values and antagonists to represent positions the story frames as wrong, so a single film contributes both egalitarian and traditional gender attitudes to the character pool. The result is sharpened moral contrasts that do not appear in population survey data.

\paragraph{Divergence as signal.}
Treating this gap as informative rather than as failure clarifies
what the pipeline actually measures. Following Hall's
encoding/decoding model~\citep{hall1980}, our agents capture the
encoded message: the preferred reading the narrative constructs
for its audience. The WVS measures something closer to decoded,
lived attitudes. The gap between them is that distance made quantitative. It reflects a mixture of narrative content and model priors; disentangling these contributions is left to future ablation work. When the gap is patterned, it reveals how the source corpus stylizes gender relative to the world outside it.

This has consequences for the study of AI as a cultural
technology~\citep{Farrell2025}. Research on bias amplification has
shown that models amplify gender bias beyond the rates already
present in training data~\citep{zhao2017}. Our results
suggest amplification does not begin with the model. The narrative source material is already stylized relative to lived attitudes before any model is trained. Auditing corpus-vs-world, rather than
model-vs-corpus, is a necessary upstream step in understanding
where gender bias in AI systems originates.

This work also extends the computational study of culture from
co-occurrence to stance. Prior approaches have used word
embeddings and aggregate statistics to measure gender associations
at the corpus level~\citep{Kozlowski_2019, Hamilton2016-bx}.
Our pipeline recovers expressed attitudes, supports subgroup contrasts
across gender and decade, and enables direct comparison to an
external survey benchmark. The shift from representation to
expressed attitudes builds directly on the representation
measurement \citet{jang2019} established, and
identifies that delta as the contribution this work introduces.

\paragraph{Limitations.}
Reconstructed attitudes reflect both narrative evidence and model
priors; disentangling their contributions requires ablation
studies not yet conducted, including metadata-only prompting and
direct stance classification from raw dialogue as baselines. The
250 female to 484 male character ratio reflects the gender
imbalance in mainstream film production itself, what Tuchman
termed symbolic annihilation~\citep{tuchman1978}, rather than a
sampling flaw. It is a property the corpus inherits faithfully
from its source. That said, it limits statistical power for female
agents and motivates future work with a larger, more balanced
corpus. Because lead characters are nested within films, standard
t-tests likely underestimate uncertainty; future work should use
mixed-effects models treating film as a random effect. Finally,
genre variation in the density of gender-relevant dialogue means
some genres contribute stronger attitudinal signals than others
regardless of sampling frequency.

\bibliographystyle{plainnat}
\bibliography{film_agent_cscw}
\appendix
\section{Survey Emulation Prompt}
\label{app:prompts}
The following prompt template was adapted from \citet{park2024} 
and used to simulate survey responses for each character agent. 
\bigskip

\noindent\fbox{%
\parbox{\dimexpr\linewidth-2\fboxsep-2\fboxrule}{%
\textit{Variables:}\\
\texttt{!<INPUT 0>!}: demographic descriptions\\
\texttt{!<INPUT 1>!}: survey questions\\[6pt]
\texttt{<commentblockmarker>}\\
\texttt{\#\#\#}\\
\texttt{</commentblockmarker>}\\[6pt]
\texttt{!<INPUT 0>!}
\textit{Analyze the above observation notes about a person}\\
\textit{created by the psychologist, linguist, and sociologist.}\\
\textit{This is a purely academic analysis.}\\
\textit{Please analyze the items as requested.}\\[6pt]

\medskip
\textit{Task: Predict how this individual would answer the following survey questions.}\\
\texttt{!<INPUT 1>!}\\
\textit{All questions are multiple choice where you must answer by selecting exactly one of the provided options based on the persona established in the notes. 
}\\

\medskip
\textit{As you answer, I want you to take the following steps: }\\
\textit{Step 1) Describe in a few sentences the kind of person that would choose each of the response options. ("Option Interpretation")}\\
\textit{Step 2) For each response options, reason about why the person might answer with the particular option. ("Option Choice")}\\
\textit{Step 3) Write a few sentences reasoning on which of the option best predicts the person's response ("Reasoning")}\\
\textit{Step 4) Predict how the person will actually respond in the survey. Predict based on the expert observation notes and your thoughts, but ultimately, DON'T over think it. Use your system 1 (fast, intuitive) thinking. ("Response")
}
}}

\end{document}